\documentclass[useAMS,usenatbib]{mn2e}
\usepackage{amsmath,amstext,amsgen,amsbsy,amsopn,amsfonts,theorem}
\usepackage{natbib}

\usepackage{graphicx}
\usepackage{xspace}
\usepackage{amsmath}
\usepackage{hyperref}
\usepackage{framed}
\usepackage{txfonts}
\usepackage{epstopdf}
\usepackage{gensymb}

\usepackage[normalem]{ulem}

\makeatletter
\def\mr@ignsp#1 {\ifx\:#1\@empty\else #1\expandafter\mr@ignsp\fi}%
\newcommand{\multiref}[1]{\begingroup
\xdef\mr@no@sparg{\expandafter\mr@ignsp#1 \: }%
\def\mr@comma{}%
\@for\mr@refs:=\mr@no@sparg\do{\mr@comma\def\mr@comma{,}\ref{\mr@refs}}%
\endgroup}
\makeatother

\newcommand{\hypref}[2]{\ifx\href\asklfhas #2\else\href{#1}{#2}\fi}
\newcommand{\Secref}[1]{Section~\multiref{#1}}
\newcommand{\secref}[1]{Sec.~\multiref{#1}}

\newcommand{\Figref}[1]{Figure~\multiref{#1}}
\newcommand{\figref}[1]{Fig.~\multiref{#1}}
\renewcommand{\eqref}[1]{(\multiref{#1})}

\newcommand{\eq}[1]{\begin{align}#1\end{align}}

\newcommand{\nln}{\nonumber\\}

\title{Forensics of Subhalo-Stream Encounters: The Three Phases of Gap Growth}

\author[D. Erkal \& V. Belokurov] {Denis
  Erkal$^1$\thanks{derkal@ast.cam.ac.uk} \& Vasily Belokurov$^1$
  \\ $^1$Institute of Astronomy, Madingley Road, Cambridge, CB3 0HA,
  UK}

\begin{document}

\label{firstpage}

\maketitle

\begin{abstract}
There is hope to discover dark matter subhalos free of stars
(predicted by the current theory of structure formation) by observing
gaps they produce in tidal streams. In fact, this is the most
promising technique for dark substructure detection and
characterization as such gaps grow with time, magnifying small perturbations into clear signatures observable by ongoing and planned Galaxy surveys. To facilitate such
future inference, we develop a comprehensive framework for studies of
the growth of the stream density perturbations. Starting with simple
assumptions and restricting to streams on circular orbits, we derive analytic formulae that describe the evolution
of all gap properties (size, density contrast etc) at all times. We
uncover complex, previously unnoticed behavior, with the stream
initially forming a density enhancement near the subhalo impact
point. Shortly after, a gap forms due to the relative change in period
induced by the subhalo's passage. There is an intermediate regime
where the gap grows linearly in time. At late times, the particles in
the stream overtake each other, forming caustics, and the gap grows
like $\sqrt{t}$. In addition to the secular growth, we find that the
gap oscillates as it grows due to epicyclic motion. We compare this
analytic model to N-body simulations and find an impressive level of
agreement. Importantly, when analyzing the observation of a single gap
we find a large degeneracy between the subhalo mass, the impact
geometry and kinematics, the host potential and the time since flyby.
\end{abstract}

\begin{keywords}
 cosmology: theory - dark matter - galaxies: haloes - galaxies: kinematics and dynamics - galaxies: structure
\end{keywords}

\section{Introduction}

Let us recall one strong and imminently testable prediction of the
modern Cosmology: in the early Universe, Dark Matter (DM) starts
collapsing first and ends up arranging itself into a hierarchy of
dense clumps of all sizes \citep[e.g.][]{White1978}. For example, by
redshift $z=0$, a DM halo with a Milky Way mass is anticipated to
contain hundreds of thousands of subhalos \citep[e.g.][]{Diemand2008,
  springel_et_al_2008}, some as massive as $10^9 M_{\odot}$, but the
majority too insignificant to kick-start star-formation, and, hence,
completely devoid of light. Nonetheless, detecting these dark halos
through their gravitational effects is feasible with existing
technology and quantifying their abundance will shed light on the
nature of Dark Matter.

Two promising experimental setups have been put forward, both to do
with the minuscule perturbations the dark substructure inflicts on
test particle orbits in the gravitational potential in question. In
one case, the role of such test particles is played by photons
traveling in the density field of a massive galaxy acting as a
gravitational lens. Intervening dark substructure then would either
cause flux anomalies in the lensed images if the source is a quasar
\citep[e.g.][]{Dalal2002} or send ripples through the lensed arcs if
the source is extended \citep[e.g.][]{Vegetti2010}. Alternatively,
Galactic stellar streams can be used as bundles of test particles to
probe the lumpiness of DM distribution.  During close flybys, the
invisible subhalos ought to ruffle the orbits of stars in the stream,
imprinting characteristic small-scale features in their density
profile. With time, such perturbation will grow, revealing a sizeable
density gap. There exists, therefore, a crucial difference between the
two experiments: the time dependence of the stream gap growth spells
out increased detectability of the DM substructure.

Evidently, if the observations of the tidal streams are to be used to
infer the mass function of the DM subhalos, it is important to
understand the time evolution of the induced density fluctuations.
However, the idea of the halo-stream interaction is relatively new
\citep[e.g.][]{ibata_et_al_2002,johnston_et_al_2002}, and, while the
overall picture has been painted with help of numerical simulations
\citep[e.g.][]{siegal_valluri_2008,carlberg_2009, yoon_etal_2011,
  carlberg_2012}, the stream dynamics due to flybys has remained
unexplained until recently when \cite{carlberg_2013} laid down the basic
equations governing the stream gap formation. In this work, we will
follow a similar strategy and consider the gaps created in a stream on
a circular orbit around an arbitrary spherical potential.

Taking advantage of this stripped-down approach, we can develop an
in-depth insight into the complex metamorphosis of the stream density
fluctuations created during encounters with dark halos. We show that,
despite the rich dynamics that ensues, many properties of the stream
gaps (e.g. gap size and density in the center of the gap) can be
solved for analytically. More generally, it is actually possible to
write a parametric function for the density profile of the stream at
all times. Importantly, our model is shown to accurately describe the
behaviour of realistic tidal streams generated in N-body simulations.

Observationally, impressive progress has been made recently in both
detecting cold stellar streams in the Galaxy
\citep[e.g.][]{pal5disc,ngc5466disc,gd1disc,bonacadisc,atlasdisc,bernard_2014} as
well as quantifying the presence of the density gaps in some of them
\citep{carlberg_pal5_2012, carlberg_gd1_2013}. Interpreting these
observations, the intuition established so far utters that the gap
size encodes predominantly the mass of the dark subhalo which wreaked
the damage \citep[e.g.][]{yoon_etal_2011}. Our analysis demonstrates
that such portrayal of the results of the halo-stream interaction is,
unfortunately, too optimistic. The inference based on the gap size
alone appears to be deeply degenerate as it is controlled by several
poorly constrained variables. As we elucidate, it is possible to
produce the same size gap in a stream by altering the dark halo mass,
the underlying host potential, the parameters of the impact, or simply
by observing the stream at a different epoch.

Fortunately, the dynamical age of the stream gap can be gleaned from
the details of the density profile in its vicinity. This is because
the gap growth proceeds in a particular sequence of phases, each
described by a specific density contrast (and its temporal evolution),
the onset timescale, and the rate of gap growth. For each of the three
phases of the stream gap growth, the compression, the expansion, and
the caustic phase, our paper provides the corresponding analytic
formulae. We, therefore, build a clear and comprehensive framework
which can be used to decipher the dark halo ballistics.

This paper is organized as follows. In
\Secref{sec:qualitative_derivation}, we begin with a qualitative
description of how stream gaps grow. We follow this with a rigorous
derivation in \Secref{sec:rigorous_derivation}. In \Secref{sec:sims},
we compare this model with idealistic N-body simulations of streams on
circular orbits, as well as a realistic N-body simulation with a
stream generated by tidally disrupting a globular cluster. In
\Secref{sec:physical_params}, we examine the degeneracy in extracting
physical parameters from gap profiles. In \Secref{sec:discussion} we
discuss how the results can be generalized and how these results can
be used to shed light on the results of previous works. Finally, we
conclude in \Secref{sec:conclusion}.

\begin{figure}
\centering \includegraphics[width=0.45\textwidth]{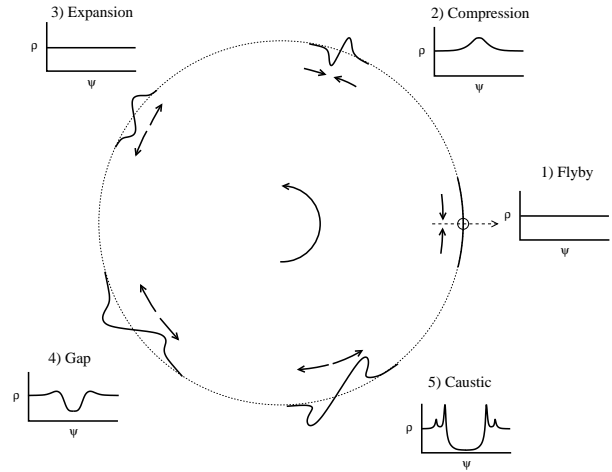}
\caption{A cartoon of gap formation and evolution. The dotted line
  delineates the orbit of the stream and the black lines show a
  segment of the stream near the point of closest approach. The graphs
  show the density along the stream for an observer in the center of
  the galaxy where $\psi$ is the angle on the sky. The arrow in the
  center shows the orbital direction of the stream and the arrows near
  the stream show whether the stream is compressing or
  expanding. Before impact, the stream has a uniform density
  (1). Shortly after, the stream is compressed since the subhalo kicks
  the particles towards the point of closest approach (2). The kicks
  also change the orbital period of particles in the stream: particles
  kicked along their orbital direction have a longer period which will
  cause them to fall behind the impact point and vice-versa. As a
  result, the compression is reversed (3) and eventually a gap forms
  (4). This expansion continues and at late times the stream particles
  overtake each other, forming caustics (5). See \protect\figref{fig:kepler_caustics}, 
  \protect\figref{fig:all_evolution}, and
  \protect\figref{fig:live_evolution} for examples of the same
  behavior in N-body simulations.}
\label{fig:gap_evolution}
\end{figure}

\section{Qualitative Explanation}  \label{sec:qualitative_derivation}

Before we present a rigorous derivation of how stellar density gaps
evolve in the toy stream model in \Secref{sec:rigorous_derivation},
let us first give a simple, intuitive explanation. For guidance, a
visual summary of the important stages of the process is also
presented in \Figref{fig:gap_evolution}.

Let us start with an unperturbed stellar stream on a circular orbit
around an arbitrary spherical potential. By restricting the analysis
to this simple case, we will be able to solve the gap growth
analytically. Bear in mind, however, that the qualitative picture
presented in this work is quite general and will hold for realistic
streams, i.e. those that have a distribution of energy and angular
momenta in their debris (see \secref{sec:realistic_sim}), as well as
for eccentric orbits (see \secref{sec:discussion}).

As a subhalo passes near (or through) the stream, its main effect is
to pull stream particles towards the point of closest approach (1 of
\figref{fig:gap_evolution}). For a wide range of encounters of
interest, these subhalo tugs are instantaneous as compared to the
stream's orbital timescale, and therefore the application of the impulse
approximation is justified. The kicks imposed by a massive perturber can
be decomposed into three components: perpendicular to the stream's
orbital plane, along the radial direction from the host, and along the
orbit. Kicks perpendicular to the orbital plane tilt the plane
slightly which causes particles to oscillate with respect to the
original plane. Radial kicks rotate the orbit in the orbital plane
which causes the density in the stream to oscillate but does not
appear to lead to any secular gap growth. Kicks along the orbit have
the biggest effect since they impart the largest change in the kinetic
energy, which changes the radial extent of the orbit and hence the
orbital period. Since the orbital period is an increasing function of
radius for any potential of astrophysical interest, particles which
are kicked along their orbit have a longer period and fall behind the
impact point. Likewise, particles which receive a kick opposite to
their orbital direction have a shorter period and race ahead of the
impact point.

Having established that the main effect of the velocity change the
subhalo imparts is to kick stream particles towards the point of
closest approach, it is straightforward to understand the three phases
of gap formation. During the \textit{compression phase}, the particles
initially move towards the impact point which creates a density
enhancement (2 of \figref{fig:gap_evolution}). After roughly an
orbital period, the changes in orbital period reverse this motion and
the gap enters the \textit{expansion phase} where particles move apart
(3 of \figref{fig:gap_evolution}), eventually forming a gap (4 of
\figref{fig:gap_evolution}). Since the magnitude of the kick depends
on position along the stream, particles will start to overtake each
other at late times which will eventually lead to the \textit{caustic
  phase} with particle pile-ups forming on either edge of the gap (5
of \figref{fig:gap_evolution}). As we will see below, one of the most
important distinctions between the expansion phase and the caustic
phase is that the the gap growth slows from being linear in time to
evolving as $\sqrt{t}$.

To complement the qualitative exposition above with quantitative
analysis, we provide a roadmap of the pertinent figures and formulae. In  \Figref{fig:kepler_caustics}, 
\Figref{fig:all_evolution}, and
\Figref{fig:live_evolution} we show the density profiles of gaps in
N-body simulations which exhibit the three phases of gap formation and can
be accurately reproduced by our model. In
\Figref{fig:central_density}, we show an example of how the density in
the center of the gap evolves in all three phases.  In
\Figref{fig:gap_size}, we show an example of how the gap size evolves
for all times. In \Figref{fig:peak_density}, we show the density in
the peaks around the gap during the intermediate phase. Lastly, we
highlight some of the useful analytic results. The expression for gap
size is given by \eqref{eq:gapsize} in the expansion phase and by
\eqref{eq:gapsizelate} during the caustic phase. The expression for
the central density is given by \eqref{eq:central_density_explicit}
and the expression for the peak density is given by
\eqref{eq:peak_density}.

\section{Rigorous Derivation} \label{sec:rigorous_derivation}

Guided by the sketch of the gap growth process as presented in the
previous section, let us now develop an analytic framework for
studying the stream density evolution after an encounter with a
subhalo. The derivation can be broken down into three main
steps. First, we will use the impulse approximation to compute the
velocity kicks the subhalo imparts along the stream. Next, we will
compute the orbits which result from these velocity kicks. Finally, we
will use these orbits to construct the stream density at all times,
allowing us to examine the gap behavior in all three phases.

\subsection{Orbit perturbation under the subhalo's impulse}

\begin{figure}
\centering
\includegraphics[width=0.3\textwidth]{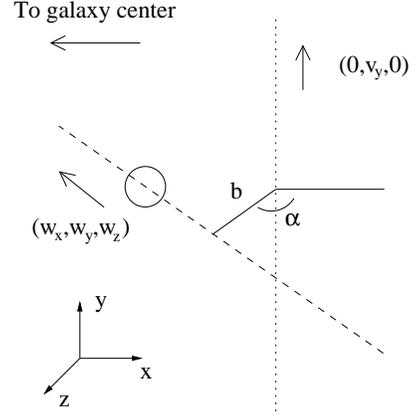}
\caption{Our axis convention looking down on the stream. The dotted line is the stream and the dashed line is the path of the subhalo. The solid lines show the impact
parameter, $b$, and its angle in the $x-z$ plane, $\alpha$. The center of the host potential is to the left and the orbit is counterclockwise.}
\label{fig:axes}
\end{figure}

The general setup for a subhalo flyby is shown in \Figref{fig:axes}
with the subhalo passing by the stream with an arbitrary geometry. We
use a similar axis convention to that in \cite{carlberg_2013} with the
stream oriented in the $y$-direction, with $x$ in the radial direction
in the host potential, and with $z$ perpendicular to the orbital
plane. The stream is moving in the positive $y$-direction in a
spherical potential $\phi(r)$, on a circular orbit with radius $r_0$,
with velocity $v_y = \sqrt{r_0 \partial_r \phi(r_0)}$. We consider a
flyby of a subhalo which is moving in an arbitrary direction with
velocity $(w_x,w_y,w_z)$ which makes a closest approach at
$(b_x,0,b_z)$, where we have chosen our coordinates and origin so the
closest approach occurs in the $x-z$ plane with the origin on the
stream. Since the impact parameter and subhalo velocity are orthogonal
at the point of closest approach, we can parameterize this point as $(b
\cos\alpha,0,b\sin\alpha)$ where $b=\sqrt{b_x^2+b_z^2}$, and the
subhalo velocity at closest approach as $(-w_\perp
\sin\alpha,w_y,w_\perp \cos\alpha)$, where $w_\perp =
\sqrt{w_x^2+w_z^2}$. Finally, we define the relative velocity between
stream and the subhalo along the stream, $w_\parallel = v_y - w_y$,
and the magnitude of the total relative velocity, $w =
\sqrt{w_\parallel^2 + w_\perp^2}$.

Let us now compute the velocity change along the stream from the passage
of a Plummer sphere with mass $M$ and scale radius $r_s$. For the
duration of the flyby, we treat the stream as a straight line,
translating at a constant velocity. In the limit that the velocity
change is small relative to the orbital velocity, we can use the
impulse approximation to get:
\eq{ \Delta v_x &= \int_{-\infty}^\infty a_x dt \nln
&= \int_{-\infty}^\infty \frac{G M (b_x + w_x t)dt}{\Big((y+w_\parallel t)^2 +
w_\perp^2 t^2 + b^2+ r_s^2\Big)^\frac{3}{2}} \nln
&= \frac{2GM \Big( b w^2 \cos\alpha + y w_\perp w_\parallel \sin\alpha \Big)}{w
\Big( (b^2+r_s^2) w^2 + w_\perp^2 y^2\Big)} . \label{eq:deltavx} }
Likewise we can compute the other two components of the velocity
change, $\Delta v_y$ and $\Delta v_z$:
\eq{ \Delta v_y = -\frac{2GM w_\perp^2 y}{w \Big( (b^2+r_s^2) w^2 + w_\perp^2
y^2\Big)} ,\label{eq:deltavy}}

\eq{ \Delta v_z = \frac{2 G M \Big(b w^2 \sin\alpha - y w_\perp w_\parallel \cos
\alpha\Big)}{w \Big( (b^2+r_s^2) w^2 + w_\perp^2 y^2\Big)} .\label{eq:deltavz} }

\begin{figure}
\centering
\includegraphics[width=0.5\textwidth]{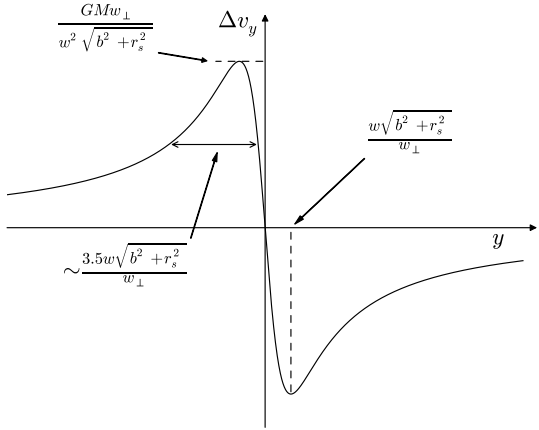}
\caption{A schematic plot of the velocity kick along the stream, $\Delta v_y$, versus position along the stream, $y$. We have marked the distance of the maximum velocity kick from the point of closest approach, the size of the maximum
velocity kick, and the width at half maximum. As we discuss in the text, the features of this relation, and the resulting orbital motion, give rise to the rich dynamics of gap formation.}
\label{fig:delta_vy_vs_y}
\end{figure}

In \Figref{fig:delta_vy_vs_y} we show a schematic plot of the velocity kick along the stream, $\Delta v_y$, versus distance from the point of closest approach. As we will see below, the features of this relation, along with the resulting orbital motion, give rise to the rich dynamics of gap evolution. 

Note that our assumption that the stream can be treated as a straight
line implies that the region over which the velocity kick occurs,
$\approx \frac{w}{w_\perp} \sqrt{b^2 + r_s^2}$, is much smaller than
the radius of the orbit, $r_0$. Furthermore, the assumption that the
stream is translating at a constant velocity implies that the duration
of the impact, $\approx \sqrt{b^2+r_s^2}/w_\perp$, is much shorter
than the orbital time, $r_0/v_y$. Therefore, we can only use these
results i) for the substructure flybys reasonably close to the
stream, ii) for the perturbers which are significantly smaller than
the stream's orbital radius, and iii) for the perturbers moving
sufficiently fast towards the stream, i.e. $\frac{w}{w_\perp}
\frac{\sqrt{b^2+r_s^2}}{r_0} \ll 1$ and $\frac{v_y}{w_\perp}
\frac{\sqrt{b^2+r_s^2}}{r_0} \ll 1$. Also note that these expressions
for the velocity kicks are similar to those that appear in
\cite{yoon_etal_2011} and \cite{carlberg_2013} due to the similarity
of the force from a Plummer sphere with that of a point mass at a
given impact parameter.

Now that we have the amplitude of the kick in each direction $(\Delta
v_x, \Delta v_y, \Delta v_z)$, we can compute the resulting orbit of
each particle along the stream. In what follows, we carry out the
analysis at leading order in $\frac{\Delta v}{v_y}$ and ignore terms
that are $\mathcal{O}(\frac{\Delta v^2}{v_y^2})$. After the kick, each
particle finds itself in a new orbital plane defined by the angular
momentum:

\eq{ L_x &= 0 , \nln
L_y &= - r_0 \Delta v_z ,\nln
L_z &= r_0 v_y + r_0 \Delta v_y . \label{eq:L} }

\noindent This new orbital plane is rotated in the $y-z$ plane in the
positive $x$ direction by $\frac{\Delta v_z}{v_y}$. If we rotate our
coordinates to align with this plane we find that the size of the
velocity kick in the new $x$ and $y$ direction is unchanged at leading
order in $\frac{\Delta v_{y}}{v_y}$ and $\frac{\Delta v_z}{v_y}$. As a
result, the kick in the $z$ direction tilts the orbital plane but
otherwise leaves the orbit unchanged. This tilt varies along the
stream and causes stream particles to oscillate in the $z$ direction
with an amplitude of $r_0 \frac{\Delta v_z}{v_y}$ with respect to the
original plane. This small oscillation is in contrast to the secular
growth of the stream gap along the orbit which we will see below. For
the rest of the analysis, we will only consider the kicks in the $x$
and $y$ direction. As is customary to describe the orbit, i.e. the
dependence of the particle's radius on the orbital phase, $\theta$, we
switch variables to $r=\frac{1}{u}$:
\eq{ \frac{d^2 u}{d\theta^2} + u = - \frac{1}{L_z^2} \partial_u \phi .}
This expression is then expanded to leading order around the original
orbit, $u = u_0 + \Delta u$, with $u_0 = \frac{1}{r_0}$, taking first
order expansions of the potential and $L_z$, to get
\eq{ \frac{d^2 \Delta u}{d\theta^2} + \gamma^2 \Delta u  = -2u_0 \frac{\Delta v_y}{v_y} , }
where $\gamma^2 =  1 + \frac{u_0^2}{v_y^2} \partial_u^2
\phi(u_0^{-1})$. The solution to the above equation is
\eq{ \Delta u = -\frac{2 u_0 \Delta v_y}{v_y } \frac{\Big( 1 -
\cos\gamma \theta \Big)}{\gamma^2} - \frac{ u_0 \Delta v_x}{v_y} \frac{\sin \gamma
\theta}{\gamma}, \label{eq:deltau}}
where we have imposed the conditions $\Delta u(0) = 0$ and
$\partial_\theta u(0) = -u_0 \frac{\Delta v_x}{v_y}$ since the stream
particle was initially on a circular orbit and received a velocity
kick, $\Delta v_x$, in the radial direction. Re-writing
\eqref{eq:deltau} in terms of $r = r_0 + \Delta r$, and expanding at
leading order in $\frac{\Delta r}{r_0}$, we get
\eq{ \Delta r = \frac{2 r_0 \Delta v_y}{v_y} \frac{\Big(1 - \cos \gamma
\theta \Big)}{\gamma^2} + \frac{r_0 \Delta v_x}{v_y} \frac{\sin \gamma \theta}{\gamma},
\label{eq:deltar} }
where we can re-write $\gamma$ in terms of $r$:

\eq{ \gamma^2 = 3 + \frac{r_0^2}{v_y^2} \partial_r^2 \phi(r_0) \label{eq:gamma} .}
Now that we know how the radius evolves after the perturbation, we can
determine the particle's angular velocity using conservation of angular
momentum:
\eq{ L_z = r^2 \dot{\theta} .}
After the impact, the angular momentum is given in \eqref{eq:L}, resulting in an angular rate of
\eq{ \dot{\theta} &= \frac{v_y}{r_0}\Big(1 + \frac{\Delta v_y}{v_y}\Big)\Big(1 +
\frac{\Delta r}{r_0}\Big)^{-2} , \nln 
&\approx \frac{v_y}{r_0}\Big(1 + \frac{\Delta v_y}{v_y} - 2 \frac{\Delta
r}{r_0}\Big). \label{eq:thetadot} }
This equation highlights the effect of the change in velocity as well
as the change in radius on the angular velocity of the orbit. Note
that if only the effect of the change in velocity is considered, no
gaps will be produced as such perturbation leads only to a density
enhancement since the $\Delta v_y$ term kicks particles towards
$y=0$. Next we can use the expression for $\Delta r$ from
\eqref{eq:deltar} in \eqref{eq:thetadot} to obtain
\eq{ \dot{\theta} = \frac{v_y}{r_0} \Big( 1 - \frac{\Delta
v_y}{v_y}\frac{4-\gamma^2}{\gamma^2} + 4 \frac{\Delta v_y}{v_y}\frac{\cos \gamma
\theta}{\gamma^2} - 2 \frac{\Delta v_x}{v_y} \frac{\sin
\gamma\theta}{\gamma}\Big). 
\label{eq:theta_almostthere}}
Finally, the orbital equation for $\theta(t)$ at leading order in
$\frac{\Delta v}{v_y}$ can be derived by re-arranging and integrating
\eqref{eq:theta_almostthere}
\eq{ \theta(t) \Big( 1 + \frac{\Delta v_y}{v_y} \frac{4 - \gamma^2}{\gamma^2}\Big)
- 4\frac{\Delta v_y}{v_y} \frac{\sin\gamma \theta(t)}{\gamma^3} + 2 \frac{\Delta
v_x}{v_y} \frac{\Big(1-\cos \gamma\theta(t)\Big)}{\gamma^2} = \frac{v_y}{r_0}t .
\label{eq:thetaoft_numerical}}
This is a transcendental equation which can be solved numerically to
give $\theta(t)$. However, since this analysis is at leading order in
$\frac{\Delta v}{v_y}$, we can approximately solve this by switching
variables to $\Delta \theta(t) = \theta(t)-\frac{v_y t}{r_0}$, and
expand at leading order in $\frac{\Delta \theta(t)}{\frac{v_y
    t}{r_0}}$, to get
\eq{ \Delta \theta(t) = - \frac{\Delta v_y t}{r_0}  \frac{4-\gamma^2}{\gamma^2} + 
\frac{4 \Delta v_y}{v_y} \frac{\sin \big(\frac{\gamma  v_yt
}{r_0}\big)}{\gamma^3} 
- \frac{2 \Delta v_x}{v_y} \frac{\Big( 1 - \cos \big(\frac{\gamma v_y
t}{r_0}\big)\Big)}{\gamma^2} ,\label{eq:thetaoft} }
which is valid as long as $\Delta \theta \ll \frac{v_y t}{r_0}$ and $\Delta \theta \ll \frac{\pi}{\gamma}$. 

\subsection{Stream evolution after the subhalo flyby}

With the analytic solution given in \eqref{eq:thetaoft}, or a numerical
solution to \eqref{eq:thetaoft_numerical}, we have a map from the
positions of particles in the stream at impact to their positions at
any later time. If we consider a particle initially at position $y$
relative to the impact point, i.e. at an angle $\psi_0 =
\frac{y}{r_0}$ relative to the point of closest approach, the position
of this particle at a later time, in coordinates which rotate with the
unperturbed stream, is given by
\eq{ \psi(\psi_0,t) = \psi_0 + \Delta \theta(\psi_0,t), \label{eq:psioft} }
where we have added the label $\psi_0$ to both $\psi$ and $\Delta
\theta$ since the velocity kick from the subhalo depends on the
initial position along the stream. In the rest of the work we will
drop the $\psi_0$ argument in $\psi(t)$ to simplify the
notation. Plugging the expression for $\Delta \theta(t)$ from
\eqref{eq:thetaoft} into \eqref{eq:psioft} gives us an analytic
formula for $\psi(t)$ in terms of the velocity kicks, $\Delta v_x,
\Delta v_y$. Using the velocity kick amplitudes from \eqref{eq:deltavx}
and \eqref{eq:deltavy} we find

\eq{ \psi(t) = \psi_0 + \frac{f\psi_0 - g}{\psi_0^2 + B^2} , \label{eq:psioftfg}
}
where
\eq{ f = \frac{4-\gamma^2}{\gamma^2} \frac{t}{\tau} - \frac{4
\sin\big(\frac{\gamma v_y t}{r_0} \big)}{\gamma^3} \frac{r_0 }{v_y \tau} 
- \frac{2 \Big( 1 - \cos\big( \frac{\gamma v_y t}{r_0}\big) \Big)}{\gamma^2}
\frac{w_\parallel}{w_\perp} \frac{r_0 }{v_y \tau} \sin \alpha , \label{eq:f} }
\eq{ g = \frac{2 \Big( 1 - \cos\big( \frac{\gamma v_y t}{r_0}\big)
\Big)}{\gamma^2} \frac{b w^2 \cos\alpha}{r_0 w_\perp^2} \frac{r_0 }{v_y \tau}, \label{eq:g} }
\eq{ B^2 = \frac{b^2 + r_s^2}{r_0^2} \frac{w^2}{w_\perp^2}, \label{eq:B} }
and the timescale $\tau$ is given by
\eq{ \tau = \frac{w r_0^2}{2 GM} .}
Note that $f, g, B, $ and $\tau$ are independent of $\psi_0$ so
although these formulae appear complicated, the map between $\psi_0$
and $\psi(t)$, \eqref{eq:psioftfg}, is quite simple in terms of
$\psi_0$. In the work below, we will do many expansions at late
time where $f$ and $g$ are dominated by the leading term in $f$, so we
also define
\eq{ f_L = \frac{4-\gamma^2}{\gamma^2}\frac{t}{\tau} , }
which is the leading order behavior of $f$ at late times.

The map between $\psi_0$ and $\psi(t)$, \eqref{eq:psioft}, allows us to immediately
compute the stream density at time $t$. If the map is single valued,
the density at $\psi = \psi(t)$ is given by
\eq{ \rho(\psi,t) = \rho_0(\psi_0) \Big| \frac{d\psi(t)}{d \psi_0} \Big|^{-1} , \label{eq:density} }
where $\rho_0(\psi_0)$ is the initial density profile of the
stream. If the map is not single valued, i.e. once
the stream particles pass each other, we must sum the right hand side
over all $\psi_0$ which map to $\psi$. Plugging the map from $\psi_0$
to $\psi(t)$, \eqref{eq:psioftfg}, into the density expression we get
\eq{ \frac{\rho(\psi,t)}{\rho_0} = \left(1 + \frac{f B^2 - f
      \psi_0^2 + 2 g\psi_0}{\big(\psi_0^2 + B^2\big)^2}\right)^{-1}
  . \label{eq:genden} }
Note that \eqref{eq:psioftfg} and \eqref{eq:genden} parametrically
define the gap profile at all times and for all impact
geometries. This gives us a functional form for the gap profile which
is very general and can be quickly computed. This can be used as a realistic
match filter to find gaps in observations \citep[i.e.][]{carlberg_pal5_2012}.

As detailed below, after the flyby the evolution of the stream
density changes behaviour several times. Let us define the timescales
which describe the phases of this metamorphosis. First, there is the
timescale for a radial oscillation which can be read off from the
expression for $\Delta \theta(t)$, \eqref{eq:thetaoft}. We will call
this the orbital timescale:
\eq{ t_{\rm orbital} \equiv \frac{r_0}{v_y \gamma}. }
Next, we have the timescale for kicked particles to reach the impact
point. Since particles at the origin receive no kick, when kicked
particles located further away along the stream reach the impact
point, they will form a caustic. This timescale is given by the
distance to the particle which receives the largest kick,
$\frac{w}{w_\perp} \sqrt{b^2+r_s^2}$, divided by the size of the kick
it receives, $\frac{GM w_\perp}{w^2 \sqrt{b^2+r_s^2}}$ (see \figref{fig:delta_vy_vs_y}). The onset of
the early caustic happens after
\eq{ t_{\rm early \, caustic} \equiv \frac{w^3}{w_\perp^2} \frac{b^2+r_s^2}{GM} . }
Lastly, we have the timescale for the particle which received the
largest kick to reverse its motion towards the impact point and reach
particles which received a negligible kick. The estimate is similar to
the case for the early caustic but now there is the added complication
of the orbital motion. This is captured in the leading term of
\eqref{eq:thetaoft} where we see that the velocity is effectively
boosted by a factor of $\frac{4-\gamma^2}{\gamma^2}$. Therefore, in
the expansion phase, the caustics will form after approximately
\eq{ t \sim \frac{\gamma^2}{4-\gamma^2} \frac{w^3}{w_\perp^2}
  \frac{b^2+r_s^2}{GM} , }
Note that the caustic timescale is derived more rigorously in
\Secref{sec:intermediate} and is given by
\eqref{eq:caustic_timescale}.

We note that while this derivation is quite general, we have made
several assumptions for the impulse approximation to hold. As we
argued in the discussion after \eqref{eq:deltavz}, our derivation of
the velocity kicks assumes that the stream can be treated as a
straight line which implies $\frac{w}{w_\perp}
\frac{\sqrt{b^2+r_s^2}}{r_0} \ll 1$. Comparing this with the
expression for $B$, \eqref{eq:B}, we see that this constraint is
equivalent to $B \ll 1$. Furthermore, if we compare the expressions
for $f$ and $g$, (\eqref{eq:f},\eqref{eq:g}), we see that $g$ is
smaller than the leading term in $f$ by a factor on the order of $B
\frac{w}{w_\perp} \frac{r_0}{v_y t}$ and smaller than the subleading
terms by a factor on the order of $B \frac{w}{w_\perp}$. In our
analysis below, we will assume that $w$ and $w_\perp$ are the same
order of magnitude so that we have $g \ll f$.

We will now analyze the consequences of these results and build a quantitative picture of the qualitative description in \Figref{fig:gap_evolution} and \Secref{sec:qualitative_derivation}.

\subsection{Early Time Behavior: Compression Phase} \label{sec:early}

At early times, $t \ll t_{\rm orbital}$, the map between $\psi_0$ and $\psi(t)$ simplifies to
\eq{ \psi(t) = \psi_0 + \frac{\Delta v_y t}{r_0} , }
i.e. particles translate along the stream at the velocity with which
they have been kicked. In terms of quantities defined above, this
becomes
\eq{ \psi(t) = \psi_0  - \frac{t}{\tau} \frac{\psi_0}{\psi_0^2 + B^2}. }
Since $\frac{d\psi(t)}{dt}$ and $\psi_0$ have opposite signs, the
stream will compress. As we have discussed earlier, this is expected
since the effect of the subhalo's passage is to pull particles towards
the point of closest approach. Note that this compression does not
depend on the details of the potential, it depends solely on the
details of the impact by the subhalo.

To characterise the stream compression, we define the center of the
perturbation as the location of the density extremum, $\frac{d
  \rho}{d\psi} = 0$, which gives $\psi_0 = 0$. Therefore, the central
density is
\eq{ \rho(0,t) = \left(1 - \frac{t}{B^2 \tau}\right)^{-1} . \label{eq:early_den}}
Thus we see that the central density increases at early times. Once
the time is on the order of the orbital time, the picture is slightly
more complicated. In general, the center of the gap is given by the
particles with $\psi_0 \approx \frac{g}{3f}$. However, since we are
only interested in the leading order behavior of the density, and we
have restricted ourselves to the $g \ll f$ regime, we will take
$\psi_0=0$ to be the center. This gives a central density of
\eq{ \rho(0,t) = \left(1+\frac{f}{B^2}\right)^{-1} . \label{eq:central_density_fB} }
Consequently, the compression reaches a maximum when $\frac{df}{dt} =
0$, i.e.
\eq{ \frac{4-\gamma^2}{\gamma^2} - \frac{4 \cos\big( \frac{\gamma v_y t}{r_0}
\big)}{\gamma^2} - \frac{2 \sin\big(\frac{\gamma v_y t}{r_0}\big)}{\gamma}
\frac{w_\parallel}{w_\perp} \sin\alpha = 0 . \label{eq:t_max_density}}
The solution to this equation will be on the order of
$\frac{r_0}{\gamma v_y}$, i.e. the compression phase lasts on the
order of a radial period, $t_{\rm orbital}$. After this time, the
particles will reverse direction due to changes in the period, leading to
the expansion phase where the density decreases, eventually forming a
gap.

\begin{figure}
\centering
\includegraphics[width=0.5\textwidth]{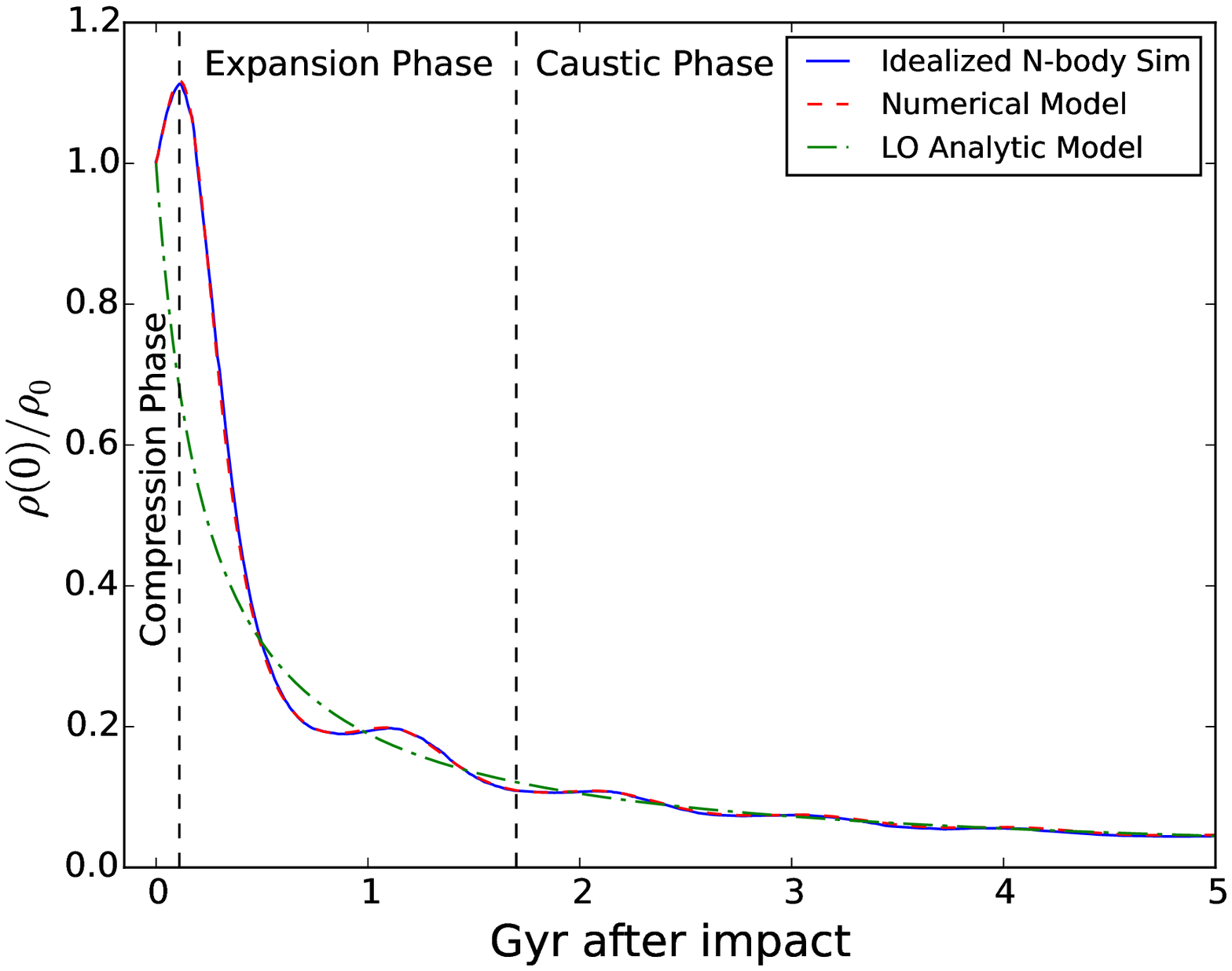}
\caption{Central density as a function of time in a Keplerian
  potential. The solid blue line shows the result of an N-body
  simulation (described in \protect\secref{sec:toy_sims}), the dashed
  red curve shows the result of the model when the central density is
  computed from a numerical solution for the density profile using
  \protect\eqref{eq:thetaoft_numerical}, and the solid green curve
  shows the result when the late time (leading order) behavior of $f$
  is used in \protect\eqref{eq:central_density_fB}. The numerical model
  reproduces the results from the N-body simulation and the leading
  order analytical model reproduces the right trends but has no
  epicyclic motion. The compression phase lasts approximately 100 Myr
  in this example, after which the expansion phase begins.}
\label{fig:central_density}
\end{figure}

In \Figref{fig:central_density}, we compare the central density in our
model to the central density in an N-body simulation of a particle
bundle and find good agreement. The N-body simulation is described in
\Secref{sec:toy_sims}. The setup is a stream-like structure on a
circular orbit with a radius of 30 kpc around a point mass with $M =
2.5 \times 10^{11} M_\odot$. The subhalo, with $M=10^7 M_\odot$ and
$r_s = $ 250 pc, directly impacts the stream with a velocity of 100
km/s perpendicular to the stream's orbital plane. In
\Figref{fig:central_density} we see that during the early part of the
compression phase, the density increases linearly as expected from
\eqref{eq:early_den}, and eventually reaches a maximum density after
approximately 100 Myr. After this, the expansion phase begins.

Apart from the density enhancement, there is one more interesting
feature during the compression phase. Since the stream particles are
initially kicked towards the point of closest approach, it is possible
that the stream particles will pass each other and form caustics
before the change in period leads to the expansion phase. This will
happen if the orbital timescale is sufficiently long compared to the
timescale for stream particles to reach the origin, i.e. the early
caustic timescale. We can determine when caustics are present when the
map between $\psi_0$ and $\psi(t)$ is multivalued, i.e. when
$\frac{d\psi(t)}{d\psi_0}=0$ has real solutions. In general, this
gives a quartic equation for which the conditions for real roots are
relatively simple but not very enlightening. However, if we restrict
to early times, $t \ll t_{\rm orbital}$, the constraint for caustics
to form simplifies to
\eq{
t_{\rm orbital} \gg B^2 \tau , }
which we can re-write as 
\eq{ t_{\rm orbital} \gg \frac{1}{2} t_{\rm early \, caustic} , }
confirming our intuition that a long orbital time is needed to form
caustics in this phase. Note that this caustic can be seen in the pole
of the density in the early phase, \eqref{eq:early_den}.

\subsection{Intermediate Time Behavior: Expansion Phase} \label{sec:intermediate}

A quick tug from the passing subhalo changes the orbital period of
particles in the stream. Even though the particles are immediately
attracted towards the impact point, with time, the orbital phase
offset due to period change accumulates and reverses the compression,
leading to a runaway gap expansion. This expansion phase continues
until eventually density caustics form as the kicked particles catch
up with more distant particles that received negligible kicks. We can
determine this time by finding the time when the map between $\psi_0$
and $\psi(t)$ becomes multivalued, i.e. when
$\frac{d\psi(t)}{d\psi_0} = 0$ has real solutions. This occurs at
$f\approx 8 B^2$ and if $g=0$, this condition is exact. This sets a
timescale for the onset of caustics, and hence the end of the
expansion phase:
\eq{ t_{\rm caustic } \approx \frac{4\gamma^2}{4-\gamma^2} \frac{w^3}{w_\perp^2} \frac{b^2+r_s^2}{GM} . \label{eq:caustic_timescale} }
Note that this is very similar to the timescale described in
\Secref{sec:rigorous_derivation}, save for the additional factor of
4. Also note that it is possible for caustics to be present during the
early part of the expansion phase if they were created in the
compression phase. These caustics will last until $f \gtrsim - B^2$,
where the condition is exact if $g=0$.

During the expansion phase, a gap forms near the point of closest
approach. \Figref{fig:gap_evolution} shows a schematic of the density profile of the
stream during the compression and expansion phase. There are three
obvious quantities of interest: the gap size, the density in the
center of the gap, and the density of the peak. We will now compute
each of these.

\begin{figure}
\centering
\includegraphics[width=0.5\textwidth]{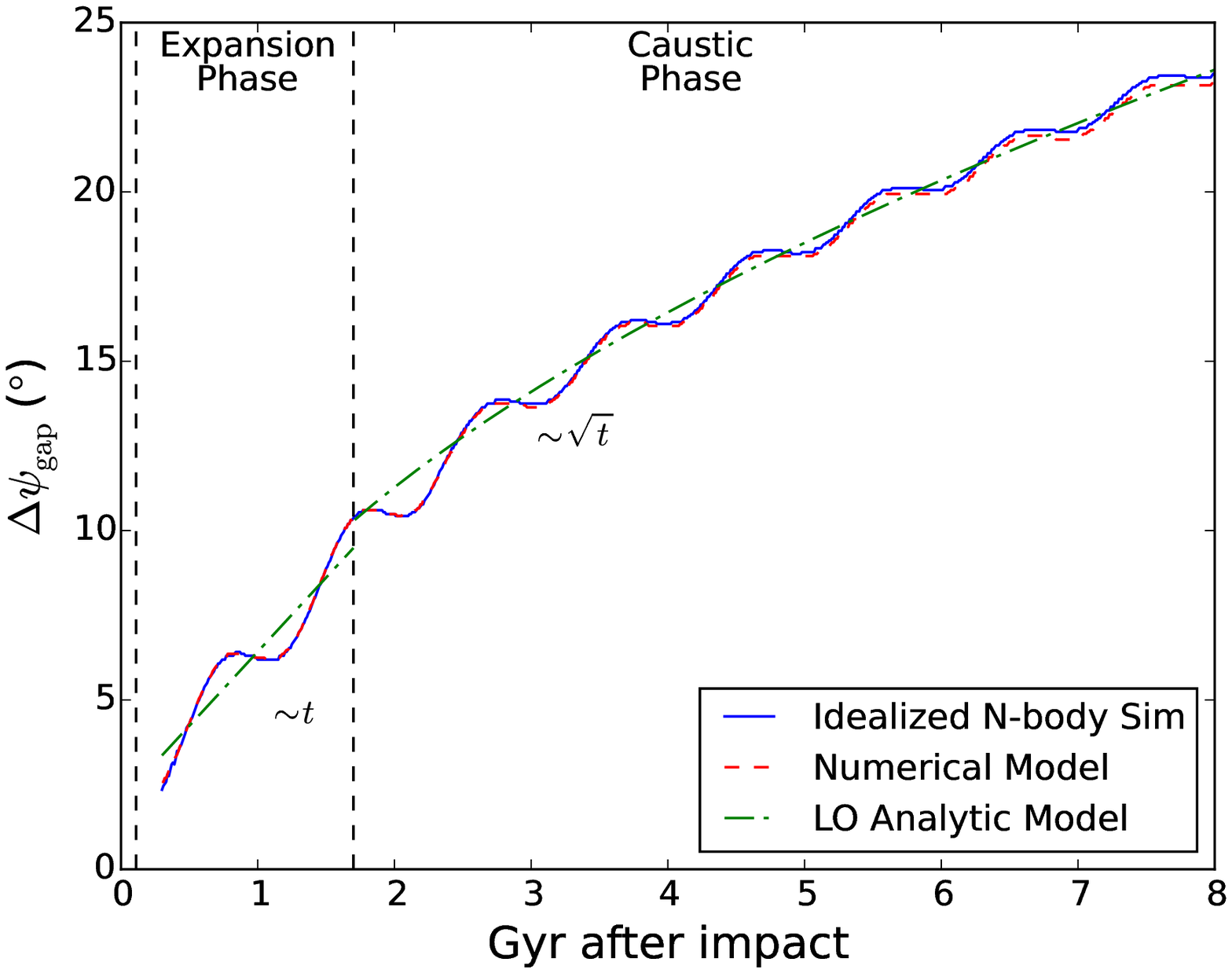}
\caption{Size of the gap as a function of time in a Keplerian
  potential. The solid blue curve shows the result of an N-body
  simulation (described in \protect\secref{sec:toy_sims}), the dashed
  red curve shows the result of the model when the gap size is
  computed from a numerical solution for the density profile using
  \protect\eqref{eq:thetaoft_numerical}, and the solid green curve
  shows the result when the late time (leading order) behavior of $f$
  is used in \protect\eqref{eq:gapsize} and
  \protect\eqref{eq:gapsizelate}. The early phase is not shown here
  since there are no gaps in that phase. The expansion phase with a
  linear growth lasts until approximately 1.7 Gyr in this
  example. After that, caustics form and the gap grows like
  $\sqrt{t}$. }
\label{fig:gap_size}
\end{figure}

After the compression stage and before the caustics form, we can write
down the density at all positions using \eqref{eq:genden}. We
define the gap size as the size of the region within which there is an
underdensity, i.e. the region within which $\frac{d\psi(t)}{d\psi_0} >
1$. The boundaries of this region are given by particles with
\eq{ \psi_0 = \frac{g}{f} \pm B \sqrt{1 + \frac{g^2}{B^2 f^2}} .}
In the limit that $f \gg g$, we can further simplify this and plugging
into the map from $\psi_0$ to $\psi(t)$, \eqref{eq:psioftfg}, we find
the leading behavior of the gap size:
\eq{ \Delta \psi_{\rm gap}(t) = 2 B + \frac{f_L}{B}
  . \label{eq:gapsize_almost} }
Note that if $g=0$, the leading order $f_L$ in this expression can be
replaced with the full $f$ expression. Plugging in for $f_L$ we get
\eq{ \Delta \psi_{\rm gap}(t) = 2  \frac{w}{w_\perp}
\frac{\sqrt{r_s^2+b^2}}{r_0} + \frac{2 GM w_\perp}{w^2
r_0 \sqrt{r_s^2+b^2}}\frac{4-\gamma^2}{\gamma^2}t  . \label{eq:gapsize} }
As is obvious from this equation, the stream gap grows linearly with time
in this phase. In addition to this linear growth, there is an
epicyclic behavior which causes the gap size to oscillate as it
grows. In \Figref{fig:gap_size} we show an example of how the gap size
grows, where the epicyclic motion is clearly visible. The simulation
setup for this example is described in \Secref{sec:early}.

The density in the center of the gap is identical to the density
during the compression phase:
\eq{ \frac{\rho(0,t)}{\rho_0} &= \left(1+\frac{f}{B^2}\right)^{-1}.}
This is the general result but we can determine the overall trend by
taking the leading term in $f$ once again to get:
\eq{ \frac{\rho(0,t)}{\rho_0} = \left(1+ \frac{4-\gamma^2}{\gamma^2}
  \frac{w_\perp^2}{w^3} \frac{2GM}{b^2+r_s^2} t\right)^{-1}
. \label{eq:central_density_explicit} }
Thus we see that the density in the center goes like $t^{-1}$ at late times. 

Finally, we can compute the position and density of the peaks around the gap. We compute these by finding the zeros of $\frac{d\rho}{d\psi}$, which gives the constraint
\eq{ 2f\psi_0(\psi_0^2 - 3B^2) - 2g(3\psi_0^2 - B^2) = 0 . }
In the limit $f \gg g$, we can neglect the second term and we see that the density peaks are at $\psi_0^2 = 3B^2$. Plugging this back into the density at late times we find
\eq{ \frac{\rho_{\rm peak}(t)}{\rho_0} = \left(1-\frac{f_L}{8B^2}\right)^{-1} . }
Note that if $g=0$, the $f_L$ in this expression can be replaced with
$f$. Plugging in the expression for $f_L$, we find
\eq{ \frac{\rho_{\rm peak}(t)}{\rho_0} =  \left(1-  \frac{1}{8} \frac{4-\gamma^2}{\gamma^2}
\frac{w_\perp^2}{w^3} \frac{2GM}{b^2+r_s^2} t\right)^{-1} . \label{eq:peak_density}}
\begin{figure}
\centering
\includegraphics[width=0.5\textwidth]{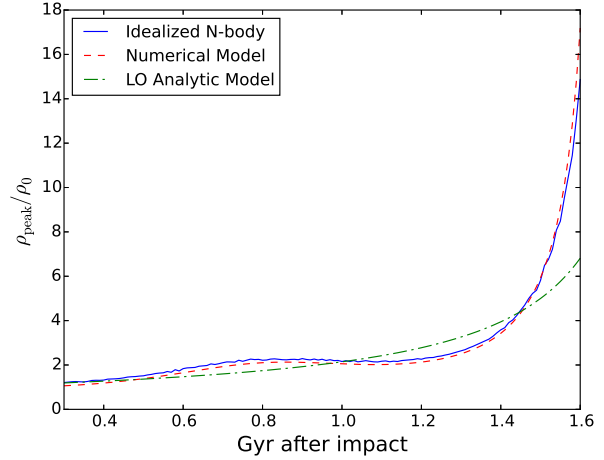}
\caption{Peak density as a function of time during the expansion
  phase. The solid blue line shows the result of an N-body simulation
  (described in \protect\secref{sec:toy_sims}), the dashed red curve
  shows the result of the model when the peak density is computed from
  a numerical solution for the density profile using
  \protect\eqref{eq:thetaoft_numerical}, and the solid green curve
  shows the result when the late time (leading order) behavior of $f$
  is used in \protect\eqref{eq:peak_density}. The early and late
  phases are not included in this plot because there is no peak in the
  early phase and because there are caustics in the late phase. Note
  that the simulation gives a consistently higher density until $\sim$
  1.4 Gyr after impact because we are taking the maximum of a density
  which is a realization with a finite number of particles.}
\label{fig:peak_density}
\end{figure}
The density diverges as we approach $t_{\rm caustic}$, heralding the
formation of caustics. We show an example of the peak density in
\Figref{fig:peak_density} which shows the asymptotic behavior. As
before, the simulation setup for this example is described in
\Secref{sec:early}.

\subsection{Late Time Behavior: Caustic Phase} \label{sec:late}

At late times, $t>t_{\rm caustic}$, stream particles overtake each
other and four caustics form, two on either side of the gap.
\Figref{fig:kepler_caustics} shows the evolution of the stream density
profile around the gap with caustics in a Keplerian potential. There are
several interesting properties we can compute: the size of the gap,
the relative strength of the caustics, the distance between the two
caustics on either side of the gap, and the characteristic width of
each caustic.

The locations of these caustics comes from determining where
$\frac{d\psi(t)}{d\psi_0} = 0$. At late times, these caustics
correspond to the particles with
\eq{ \psi_{0, \rm inner}^2 = f_L , }
and
\eq{ \psi_{0, \rm outer}^2 = B^2 , }
where the labels refer to the caustic position relative to the impact
point during the caustic phase. We can then plug these into the map
between $\psi_0$ and $\psi(t)$, \eqref{eq:psioftfg}, to determine their
positions:
\eq{ \psi_{\rm inner}(t) &= \pm 2 \sqrt{f_L}, \nln \psi_{\rm outer}(t) &=  \pm \Big(B + \frac{f_L}{2B}\Big) . }
Thus, comparing with \eqref{eq:gapsize_almost} and
\eqref{eq:gapsize} we see the outer caustic moves linearly in time and
continues at the same rate as the gap edge during the expansion
phase. In addition, there is an inner caustic which moves
proportionally to $\sqrt{t}$. As we will see below (also see
\figref{fig:kepler_caustics}), the inner caustic is both stronger and
wider than the outer caustic. Thus the inner caustic sets the gap size
in this phase:
\eq{ \Delta \psi_{\rm gap}(t) = 4 \left(\frac{4-\gamma^2}{\gamma^2} \frac{2GM}{w r_0^2}t\right)^{\frac{1}{2}}. \label{eq:gapsizelate} }
\Figref{fig:gap_size} shows the gap size evolution during the caustic
phase. We see that the model closely matches the N-body
simulation. Similarly, \Figref{fig:central_density} shows the density
in the center of the gap during the caustic phase and once again
reveals good agreement with the N-body simulation.

Another useful prediction during this phase is the relative strengths
of the inner and outer caustics which are proportional to $\Big|
\frac{d^2 \psi(t)}{d\psi_0^2}\Big|^{-\frac{1}{2}}$ evaluated at the
caustic position,
\eq{ \frac{\rho_{\rm inner}(t)}{\rho_{\rm outer}(t)} = \frac{f_L^\frac{3}{4}}{2 B^\frac{3}{2}} . }
Thus we see that the inner caustic dominates the outer caustic for $t
\gg t_{\rm caustic}$.

The distance between the inner and outer caustic is another
interesting quantity since it gives the size of the overdensity region
around the gap. This bump size is given by
\eq{ \Delta \psi_{\rm bump} = B + \frac{f_L}{2B} - 2\sqrt{f_L} . \label{eq:bump_size}}

Finally we note that these caustics can be extremely narrow. Their
characteristic widths are given by $\Big| \frac{d^2
  \psi(t)}{d\psi_0^2}\Big|^{-1}$ evaluated at the caustic. At late
times, the widths of the inner and outer caustic are given by:
\eq{ \Delta \psi_{\rm inner} \approx \frac{\sqrt{f_L}}{4} , }
\eq{ \Delta \psi_{\rm outer} \approx \frac{B^\frac{3}{2}}{2 \sqrt{f_L}} . }
Thus we see that the width of the inner caustic grows with time while
the width of the outer caustic shrinks (as illustrated in
\Figref{fig:kepler_caustics}).

\section{Comparison with Simulations} \label{sec:sims}

To demonstrate the validity of the derivation above we have carried
out several N-body simulations. These simulations were all run with
the pure N-body part of GADGET-3 which is closely related to GADGET-2
\citep{springel_2005}.

\subsection{Idealized Simulations}  \label{sec:toy_sims}

The first set of simulations we carried out are similar to those in
\cite{carlberg_2013} and mimic the setup of the derivation above. We
placed $10^6$ massless tracer particles on a short arc (0.6 radians)
on a circular orbit with $r_0 = $ 30 kpc and a circular velocity of
$v_y =$ 190 km/s. These arcs were evolved in three different
potentials: NFW \citep{nfw_1997}, Keplerian, and spherical harmonic
oscillator (SHO). The NFW has a mass of $M = 10^{12} M_\odot$, a
concentration of $c=15$, and a scale radius $R_s=14.0$ kpc. The
parameters of the Keplerian and SHO potential were chosen to have the
same circular velocity as the NFW potential at 30 kpc, resulting in a mass of
$2.5 \times 10^{11} M_\odot$ for the Keplerian potential, and a spring
constant of $k = 40$ km$^2$/s$^2$/kpc$^{2}$ for the SHO potential.

We modified GADGET to include a subhalo particle which moves at a
constant velocity and feels no force but exerts a Plummer force on the
other particles. This was done to mimic the setup of the toy model and
avoid any complications arising from orbit of the subhalo. The subhalo
particle has a velocity of $w_z = $ 100 km/s and exerts a Plummer
force with $M=10^7 M_\odot$ and $r_s =$ 250 pc. The initial conditions
for the subhalo particle and the stream particles were setup so that
the impact would occur in the middle of the stream, i.e. a direct
impact perpendicular to the stream's orbital plane.

\Figref{fig:all_evolution} shows the stream density profiles in
simulations with three different potentials at several epochs. As we
saw in the derivation above, the gap evolution in each potential is
controlled by $\gamma$, \eqref{eq:gamma}. The NFW potential has
$\gamma^2 = 2$, the Keplerian potential as $\gamma^2 = 1$, and the SHO
potential has $\gamma^2 = 4$. The streams in all three potentials show
a density enhancement at early times. Note that the compression phase
is identical for all three potentials since the early behavior is
independent of potential. The compression is followed by an expansion
phase and in the NFW and Keplerian potentials, the expansion phase
results in gap growth. As expected from the expression for the gap
size, \eqref{eq:gapsize}, the gap grows three times faster in the
Keplerian potential. Interestingly, there is no secular gap growth in
the SHO potential since the orbital period in a SHO is independent of
radius. Instead, the stream oscillates between an overdensity and an
underdensity.

\begin{figure}
\centering
\includegraphics[width=0.5\textwidth]{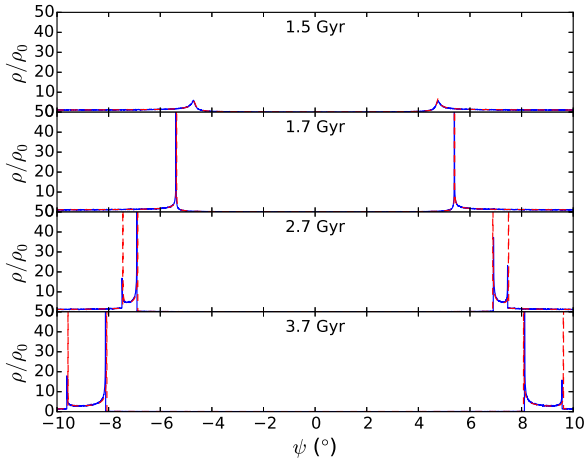}
\caption{Evolution of caustics at late time for the Keplerian
  potential. The solid blue curve shows the result of an N-body
  simulation and the dashed red curve shows the prediction from the
  model using a numerical solution of
  \protect\eqref{eq:thetaoft_numerical}. The caustics in this example
  form roughly 1.7 Gyr after impact.}
\label{fig:kepler_caustics}
\end{figure}

\begin{figure*}
\centering
\includegraphics[width=1.0\textwidth]{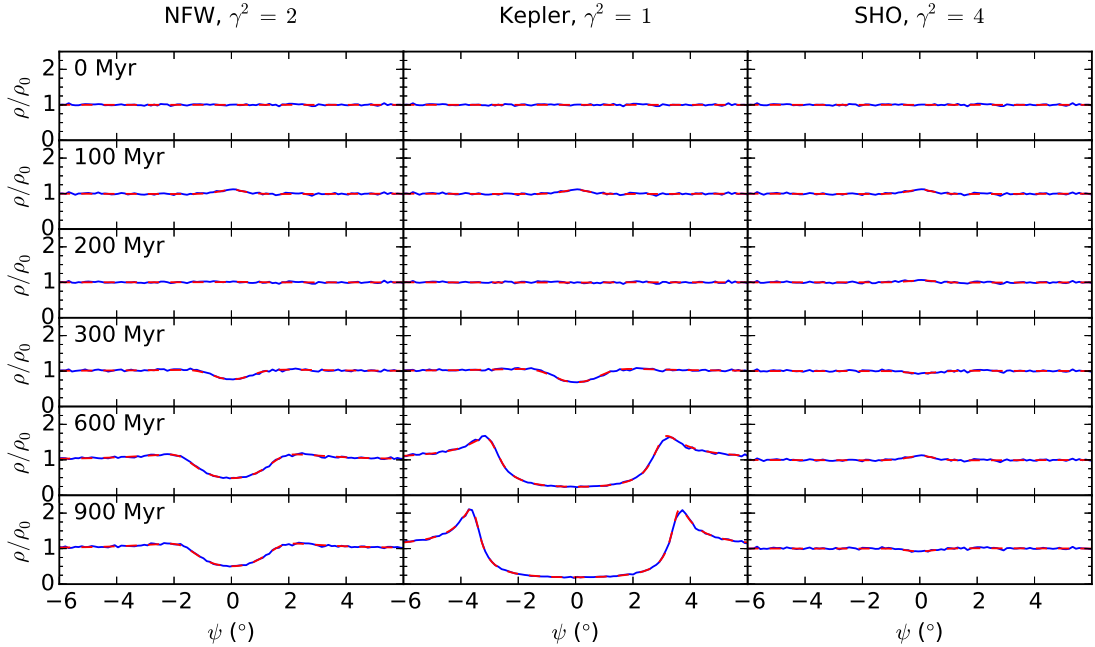}
\caption{Density profile of the stream after an impact in three
  different host potentials. The left panel is for an NFW host, the
  middle panel is for a Keplerian host, and the right panel is for a
  simple harmonic oscillator host. The solid blue curve shows the
  result of an N-body simulation and the dashed red curve shows the
  prediction from the model using a numerical solution of
  \protect\eqref{eq:thetaoft_numerical}. As described in the text, the
  gap grows three times faster in the Keplerian potential than the NFW
  potential in the expansion phase. In addition, gaps do not grow in
  the SHO potential since the period is independent of radius.}
\label{fig:all_evolution}
\end{figure*}

The caustic phase is not visible in \Figref{fig:all_evolution} but we
have shown the caustic phase for the Keplerian potential in
\Figref{fig:kepler_caustics}. We see that the model correctly predicts
the locations of the double caustics on either side of the gap, as
well as the density profile of the gap. The reason we do not show the
NFW potential in \Figref{fig:kepler_caustics} is that it will not have
entered the caustic phase by the final panel of
\Figref{fig:kepler_caustics}. As we can read off from the caustic
timescale, \eqref{eq:caustic_timescale}, the caustic timescale for the
NFW potential is three times longer than the caustic timescale for the
Keplerian potential.

Additionally, several properties of the stream gaps in the Keplerian
potential are compared against the predictions of our model. Namely,
\Figref{fig:central_density} shows the central density in the
intermediate and late phase, \Figref{fig:gap_size} gives the gap size,
and \Figref{fig:peak_density} presents the density of the peaks around
the gap in the intermediate phase. In all cases, we find good
agreement between the simulation and the model.

\subsection{Realistic Simulation} \label{sec:realistic_sim}

In the previous section, we compared the analytical model to
simulations of an interaction of a subhalo with a stream consisting of
particles on the same circular orbit. In realistic streams produced
through tidal disruption, the debris particles would have a
distribution of energies and angular momenta. Moreover, the progenitor
of the stream is constantly being stripped resulting in particles
which can potentially fill in the gap. To show that our simple model
is still useful for gaps in a realistic streams, we have carried out a
N-body simulation where we have a subhalo directly impact a tidal
stream generated by disrupting a globular cluster.

We model the globular cluster as a Plummer sphere with a mass of
$2.5\times 10^5 M_\odot$ and a scale radius of $8$ pc. It is placed on
a circular orbit of radius $10$ kpc around an NFW potential with $M =
10^{12} M_\odot$, with $c=15$ and $R_s=14.0$ kpc. The Plummer sphere
is represented with $10^6$ particles which have a smoothing length of
0.43 pc. This smoothing length is used to minimize the force errors
\citep{dehnen_2001}. The Plummer sphere is evolved for 3 Gyr and in
this time a long cold stream with a length of $\sim 180 \degree$ is
produced. As described above, we add a subhalo particle which moves
with a fixed velocity and exerts a Plummer force, with $M=10^8
M_\odot$ and a scale radius of $r_s = $ 250 pc, on all other
particles. This particle is setup to impact the center of the leading
arm of the stream, which corresponds to a radius of 9.77 kpc, at 100
km/s perpendicular to the orbital plane. We then follow the evolution
of the stream for 4 Gyr after the impact to see how the stream
evolves. Note that we have slightly modified the setup in \Secref{sec:toy_sims}, using a more massive halo on an orbit with a smaller radius, to make the gap more pronounced.

\begin{figure}
\centering
\includegraphics[width=0.5\textwidth]{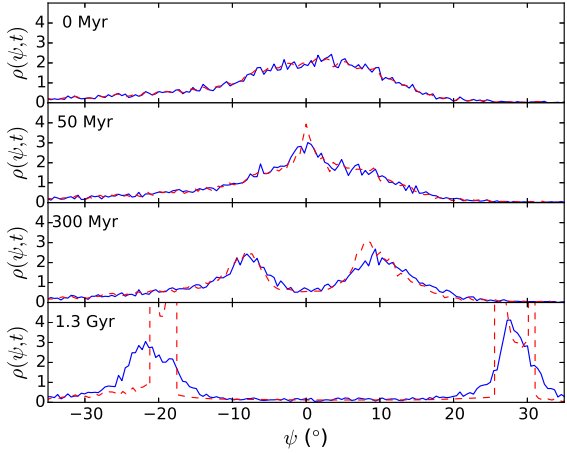}
\caption{Evolution of the stream density in an N-body simulation of a
  subhalo impact on a stream produced by disrupting a Plummer
  sphere. The solid blue curve shows the result of an N-body simulation
  and the dashed red curve shows the prediction from the model using a
  numerical solution of \protect\eqref{eq:thetaoft_numerical}.}
\label{fig:live_evolution}
\end{figure}

For our analytic model, we assume the stream particles are on a single
circular orbit with $r_0 = 9.77$ kpc and use the circular velocity at
this radius, 168.2 km/s, for the stream velocity. We also have to
account for the fact that the unperturbed stream now has a non-trivial
density profile. This is naturally included in our model since the
density is related to the initial density through
\eqref{eq:density}. Note that before the halo flyby, the stream
has developed a broad density enhancement close to the end of the stream
 (top panel of \figref{fig:live_evolution}). This is a consequence of the fact that the Plummer sphere we inserted on a circular orbit is initially out of equilibrium with the tidal field and has a stripping rate which is a decreasing function of time. This results in a peak of the density along the stream. We chose to directly impact this peak to avoid any additional confusion
in interpreting density peaks not created by the gap.

\Figref{fig:live_evolution} shows the density profile along the stream
at various times. We see the same behavior as we found in
\Secref{sec:rigorous_derivation}: there is an initial density
enhancement which gives rise to a gap with peaks around it. The
caustics which were prominent in the toy model
(\figref{fig:kepler_caustics}) are now mostly smoothed over by the
dispersion in E-L of the stream debris and by the non-trivial shape of
the initial density profile. In the lowermost panel, we see that there
are some small bumps near where the caustics should be. While these
bumps are marginally visible in many of the snapshots at the correct
location, it is unclear if we are actually resolving them.

\begin{figure}
\centering
\includegraphics[width=0.5\textwidth]{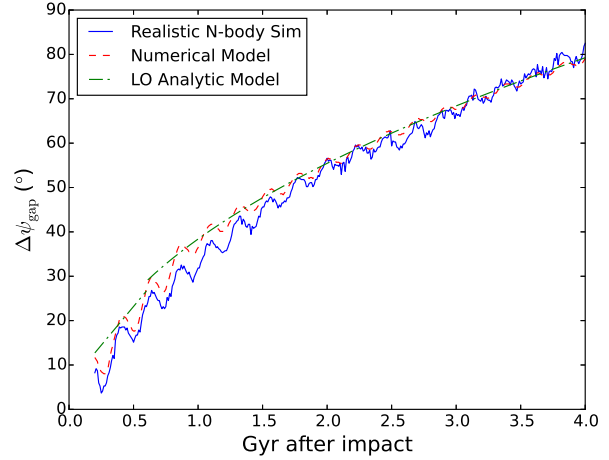}
\caption{Gap size in an N-body simulation of a disrupting Plummer
  sphere. The solid blue curve shows the result of an N-body
  simulation, the dashed red curve shows the result of the model when
  the gap size is computed from a numerical solution for the density
  profile using \protect\eqref{eq:thetaoft_numerical}, and the solid
  green curve shows the result when the late time (leading order)
  behavior of $f$ is used in \protect\eqref{eq:gapsize} and
  \protect\eqref{eq:gapsizelate}.}
\label{fig:live_gap_size}
\end{figure}

\Figref{fig:live_gap_size} illustrates how the gap size evolves and
reports a fairly good agreement between the result of the realistic
N-body simulation and our model. We see that despite the lack of the
distinctive caustic features, the gap size growth starts off linear
and then slows to be proportional to $\sqrt{t}$, in agreement with our
model. One possible reason for a small discrepancy with the N-body
result having a slightly steeper slope than our model is likely due to
the fact that the stream is not on a single circular orbit, as we have
assumed, but rather the particles sample a sequence of orbits which
are stretching away from the progenitor due to the difference in
angular momentum and hence period. This causes the unperturbed stream
to stretch out which will increase the rate of gap growth.

\begin{figure}
\centering
\includegraphics[width=0.5\textwidth]{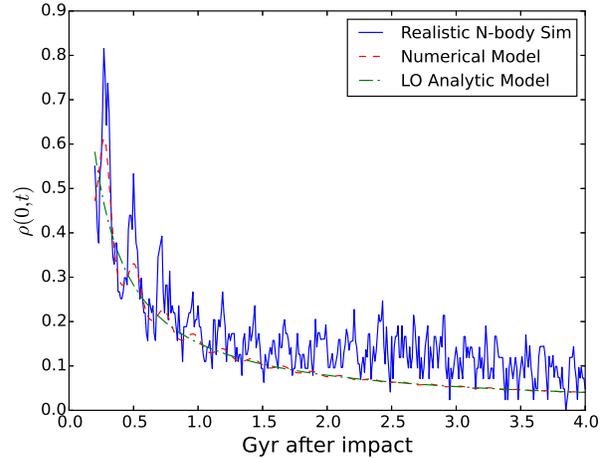}
\caption{Density in the center of the gap as a function of time. The solid
  blue curve shows the result of an N-body simulation, the dashed red
  curve shows the result of the model when the central density is
  computed from a numerical solution of the density using
  \protect\eqref{eq:thetaoft_numerical}, and the solid green curve
  shows the result when the late time behavior of $f$ is used in
  \protect\eqref{eq:early_den}. We see that our model reproduces the
  density at early times but underpredicts it at late times. This is
  likely due to stream particles filling the gap due to their
  distribution in energy and angular momentum.}
\label{fig:live_trough_density}
\end{figure}

\Figref{fig:live_trough_density} compares the density in the center of
the gap in the N-body simulation against our model. Overall, we find a
very good agreement at early times, but our model slightly
underpredicts the density at late times. This is likely due to the
stream particles filling in the gap since they have a spread in energy and angular momentum, an effect not included in our
model. Due to this dispersion, the gap
can be filled by material which is stripped from the progenitor at a
later time.

\begin{figure}
\centering
\includegraphics[width=0.5\textwidth]{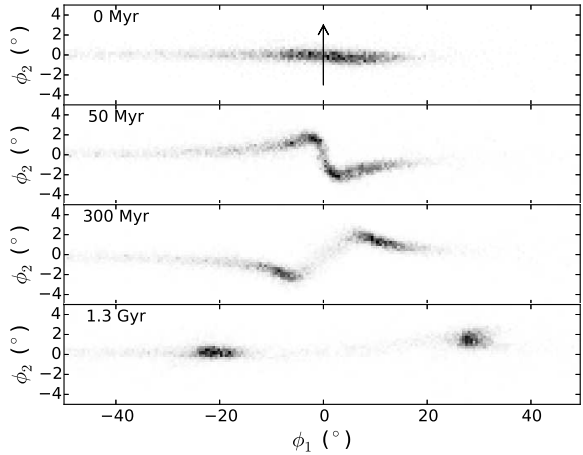}
\caption{Stream resulting from the disruption of a Plummer sphere as viewed from the center of the galaxy at different
  times. $\phi_1$ is aligned with the orbital plane of the unperturbed
  stream. The times in the top left show the time since impact. The
  arrow shows the motion of the subhalo which impacts the stream at
  $\phi_1=0$ at $t=0$. Due to the geometry of this particular
  encounter, the stream oscillates out of the orbital plane. However,
  the main effect of the subhalos passage is the formation of a gap
  which grows like $\sqrt{t}$, as in
  \protect\figref{fig:live_gap_size}.}
\label{fig:live_stream_on_sky}
\end{figure}

\Figref{fig:live_stream_on_sky} gives the on-sky picture of the stream
as viewed from the center of the galaxy. Since the subhalo is moving
perpendicular to the orbital plane, the stream particles receive a
kick in that direction which causes the stream particles to oscillate
perpendicular to the orbital plane. However, the main effect is for
the stream to stretch out along the orbital direction.

\section{Extracting Physical Parameters from Gaps} \label{sec:physical_params}

Now that we understand how the growth of stream gaps depends on the
host potential and the properties of the subhalo, let us elucidate the
inverse problem: given the gap properties, can useful constraints be
placed on the properties of the subhalo? To answer this question, we
first have to think about the observables of the gap. These
observables depend on the phase the gap is in. During the compression
phase, the only feature is the density enhancement. While this feature
potentially presents a useful constraint on the subhalo flyby
properties, it is unlikely this phase will be observable due to its
short lifetime. During the expansion phase, the simplest set of
observables would be the size of the gap, the density in the center of
the gap, and the density in the peaks around the gap. During the
caustic phase, the observables would be the size of the gap, the
relative strength of the caustics, the distance between caustics, and
the width of the caustics. Alternatively, we could attempt to fit the
gap profile with the parametric form of the density profile. We will
discuss both approaches below. For clarity of the following
discussion, we re-write $f_L$ and $B^2$ since these two parameters
control the overall behavior of observables mentioned above
\eq{ f_L = \frac{4-\gamma^2}{\gamma^2} \frac{2GM}{w r_0^2} t ,}
\eq{ B^2 = \frac{b^2+r_s^2}{r_0^2} \frac{w^2}{w_\perp^2}. }

\begin{figure}
\centering
\includegraphics[width=0.5\textwidth]{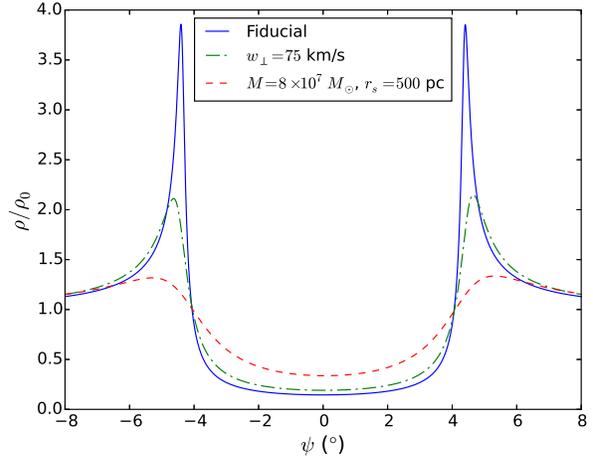}
\caption{The density profile for various flybys which have the same
gap size but different profiles. The fiducial setup is for a subhalo with a mass of $10^7 M_\odot$, a scale radius of 250 pc, and a perpendicular velocity, $w_\perp$, of 100 km/s. As discussed in the text, for a given gap size, the density contrast increases as $B$ decreases. As a result, for a given gap size, a smaller perpendicular velocity, $w_\perp$, results in a smaller density contrast. Similarly, for a given gap size, a more massive and more extended subhalo results in a smaller density contrast.}
\label{fig:gap_degeneracy}
\end{figure}

As we saw in \Secref{sec:intermediate}, in the expansion phase, the
gap size is governed by the quantities $B$ and $f_L/B$, however at
late times the second term is significantly larger than the first and
is responsible for the growth. As a result, we can think of the gap
size as controlled by the combination $f_L/B$. We also found that the
density of the peak and trough is controlled by $f_L/B^2$. Therefore,
for a given gap size, the density contrast increases as $B$
decreases. In \Figref{fig:gap_degeneracy}, we demonstrate this with an
example of three gaps which are identical in size but have different
density contrasts. The setup is identical to the setup above: the
stream is on a circular orbit with $r_0 = $ 30 kpc, around a NFW with
$M=10^{12} M_\odot$, $c=15$, and $R_s = 14.0$ kpc. The fiducial
subhalo is a Plummer sphere with $M=10^7 M_\odot$, $r_s = 250$ pc, and
$w_\perp = 100$ km/s. Note that the density profiles in \Figref{fig:gap_degeneracy} occur
at different times in the different setups since we require the gaps to have the same size. Thus, we see that by measuring the gap size and the density
contrast we can constrain $f_L$ and $B$.

In the late phase, the argument is similar, except now the gap size
goes like $\sqrt{f_L}$ and the size of the overdense region around the
gap depends on $B$ and $f_L$ as in \eqref{eq:bump_size}. As a result,
we can once again constrain $f_L$ and $B$. This argument extends to
the other properties during the caustic phase, i.e. the relative
strength of the caustics and their width, which also depend on
combinations of $f_L$ and $B$.

\subsection{Degeneracy for Single Gap}

Given that the gap size and the density contrast only depend on two
quantities, $f_L$ and $B$, we see that there will be a large
degeneracy when inferring subhalo properties. If we assume that we
know the orbital properties of the stream, i.e. $r_0$ and $v_y$, as
well as the host potential, $\gamma$, we see that the gap properties
depend on 7 quantities: $M,r_s,b,w_\perp,w_\parallel,\alpha,t$. These
7 quantities are constrained by $f_L$, $B$, and a constraint on the
density of the subhalo, $\frac{M}{r_s^3}$. Thus we are left with four
unconstrained degrees of freedom.

This picture is further complicated by the epicyclic motion which
causes the gap size and the density to oscillate
(i.e. \figref{fig:central_density}, \figref{fig:gap_size}). As a
result, when these properties are measured, there will be some
uncertainty about exactly what phase of the expansion they are in.

In the best case scenario, we could fit the stream density profile
with the parametric function for the stream density
(\eqref{eq:psioftfg} and \eqref{eq:genden}). Looking at these
equations, we see that this parametric function only depends on $f,g,$
and $B$. Therefore, even if we fit the exact stream profile, we will
be left with a three dimensional degeneracy. This means that it is not
possible to uniquely infer the properties of a subhalo from a gap
profile, even in the most optimistic case.

\subsection{Constraints from Gap Spectrum}

This gloomy prediction may improve somewhat if we instead try to model
the spectrum of gaps created by multiple encounters with subhalos. For
a statistical sample of gaps, we could use additional information to
constrain the distribution of velocities and impact parameters using
constraints on the position and velocity distributions of the subhalos
from N-body simulations. The gap spectrum would then allow us to potentially
constrain the subhalo mass function. Note that this analysis would be complicated by overlapping gaps
as well as the epicyclic overdensities expected in streams, e.g. \cite{kupper_etal_2008}. 

\section{Discussion} \label{sec:discussion}

\subsection{Generalizations}

In the work above, we have built a model for gaps formed by the flyby
of a Plummer sphere near a stream on a circular orbit. While this
model contains many simplifying assumptions, the qualitative picture
we have developed holds for more generic encounters. The three
distinct phases, as well as the transition from a gap which grows
linearly in $t$ to one which grows like $\sqrt{t}$, will be present in
the flybys of generic subhalos near streams on non-circular orbits
whose particles have a distribution of energies and angular momenta.

The generalization to different subhalo density profiles affects the
resulting velocity kicks, $(\Delta v_x,\Delta v_y,\Delta v_z)$. For a
general spherically symmetric subhalo profile, these kicks must be
evaluated numerically. However, as long as the kicks produced have the
same qualitative features as the ones generated by a Plummer sphere,
i.e. a similar shape in the plane of velocity change versus distance
from the point of closest approach (i.e. as in \figref{fig:delta_vy_vs_y}), the qualitative picture will
remain true. For example, for an NFW profile, the radial force does
not go to zero as we approach the origin. As a result, the velocity
kick for particles near the impact point will change rapidly along the
stream (i.e. making \figref{fig:delta_vy_vs_y} steeper near the origin), making it much easier for caustics to form in the compression
phase. This can be understood in terms of the early caustic timescale
which is the distance to the largest kick, divided by its
amplitude. However, the intermediate and late time behavior will still
be the same since they are due to the change in the orbital period and
the particles with the largest kick catching up to those which
received a negligible kick. Note that these differences are even
smaller for impact parameters larger than the scale radius of the
impactor since then the precise profile becomes unimportant.

The extension to eccentric orbits is non-trivial for a general
potential since the orbits are not analytic. However, the qualitative
picture in \Secref{sec:qualitative_derivation} holds for non-circular
orbits so we expect the same overall behavior seen here. For eccentric
orbits, the gap size will oscillate more dramatically as it grows due
to the difference in angular velocity from pericenter to apocenter. In
addition, the effect of the subhalo's passage will also depend on
where along the orbit the closest approach occurs. For a fixed kick
size, a kick at pericenter will have a larger effect on the kinetic
energy and hence the period compared to a kick at apocenter. However,
this simple picture is complicated by the fact that both the stream
particles and the subhalo will likely have a lower velocity at the
stream's apocenter, resulting in a larger kick at apocenter. Despite
these complications, as we show in \Secref{sec:previous_work}, the
scaling behaviors of the gap size reproduce what is seen in
cosmologically motivated suite of simulations by
\cite{yoon_etal_2011}. In addition, N-body simulations (not shown
here) of eccentric orbits around NFW potentials show the same
qualitative behavior with an overdensity at early times, leading to a
gap, and finally to caustics at late times.

The extension to streams with a distribution of energy and angular
momenta was shown in the N-body simulation in
\Secref{sec:realistic_sim} where we first generated a stream by
disrupting a Plummer sphere, and then generated a gap with the flyby
of another Plummer sphere. Despite the realistic distribution in
energy and angular momentum, the gap size growth still exhibits the
linear growth in $t$ in the expansion phase and the $\sqrt{t}$ growth
in the caustic phase (\figref{fig:live_gap_size}). In addition, there
is a density enhancement visible at early times
(\figref{fig:live_evolution}).

\subsection{Dependence on Potential} \label{sec:gap_size_pot}

To develop some intuition about how the gap size growth depends on the
potential we consider the power-law potential, $\phi = Ar^n$, which
would imply that $\gamma^2 = 2+n$. Plugging this into the expressions
for the gap size (\eqref{eq:gapsize} or \eqref{eq:gapsizelate}) we
find that the growth rate is proportional to $\frac{2-n}{2+n}$ during
the expansion phase and $\sqrt{\frac{2-n}{2+n}}$ during the caustic
phase. Therefore, as $n$ approaches -2 the gap grows faster and
faster. This follows from the fact that the effective potential,
$\phi(r) + \frac{L_z^2}{2r^2}$, expanded around the radius for a
circular orbit, becomes flatter and flatter in this limit and thus the
radial oscillations get larger. Since the period is an increasing
function of radius for potentials with $n < 2$, these radial
oscillations lead to dramatically different periods and hence a
rapidly expanding gap. As $n$ approaches 2, the gap grows more slowly
since it is approaching a spherical harmonic oscillator where the period
is independent of radius and no gap will form, as shown in
\Figref{fig:all_evolution}.

\subsection{Simplified Picture} \label{sec:simple}

In \Secref{sec:qualitative_derivation} and
\Secref{sec:rigorous_derivation} we gave a qualitative and a rigorous
derivation of how gaps grow. These results can be summarised quite
neatly. The formation of gaps is governed by three timescales: the
orbital timescale, $t_{\rm orbital}$, the early caustic timescale,
$t_{\rm early\, caustic}$, and the caustic timescale, $t_{\rm
  caustic}$. Within an orbital timescale, the stream will compress,
expand, and then begin to form a gap. If $t_{\rm early\, caustic} \ll
t_{\rm orbital}$, early caustics will form in the compression phase
and vanish before the expansion phase. Between the orbital timescale
and the caustic timescale the stream gap will grow linearly in
time. After the caustic timescale, caustics form on the leading edge
of the gap and the gap size goes like $\sqrt{t}$. In terms of $B$ and
the caustic timescale, the gap size and densities are also remarkably
simple. In the intermediate phase, the gap size is given by
\eq{
\Delta \psi_{\rm gap} = 2 B + 8 B \frac{t}{t_{\rm caustic}} ,
}
the density of the peaks around the gap is given by
\eq{
\frac{\rho_{\rm peak}(t)}{\rho_0} = \left(1- \frac{t}{t_{\rm caustic}}\right)^{-1} ,
}
and the density in the center of the gap (which holds in the
intermediate phase and the late phase) is given by
\eq{
\frac{\rho(0,t)}{\rho_0} = \left(1+ \frac{8t}{t_{\rm caustic}}\right)^{-1}.
}
During the late phase, the gap size is given by
\eq{
\Delta \psi_{\rm gap} = 8\sqrt{2} B \sqrt{\frac{t}{t_{\rm caustic}}} .
}
Thus we see that the stream properties are especially simple when
expressed in terms of $t_{\rm caustic}$ and $B$. For example, we can
immediately see that if a gap has a very small density in the center,
$\rho/\rho_0 < 0.1$, the gap is in the caustic phase and the gap size
is growing as $\sqrt{t}$.

\subsection{Comparison with Previous Work} \label{sec:previous_work}

In this work, we have extended the results of \cite{carlberg_2013} to
the formation of gaps in arbitrary host potentials and to subhalos
which are Plummer spheres. As in that work, we use the impulse
approximation to compute the kick on stream particles from the passage
of a subhalo. In \cite{carlberg_2013}, the effect on the stream is
computed analytically using guiding centers and epicyclic motion for
the case of a logarithmic potential. The results found in
\cite{carlberg_2013} match the qualitative behavior in the expansion
phase of this work with a gap size that grows linearly in
time. However, our results differ from those in \cite{carlberg_2013}
since the expression for the gap size in that work \citep[Equation 16
  of][]{carlberg_2013} has a different scaling behavior with larger
mass subhalos giving smaller gap sizes and also appears to have
typographical errors since the units are inconsistent. In addition, we
find a richer structure with three phases of gap formation and a
different gap growth at late times. Note that there are hints of the
three phases of gap formation in Figure 6 of \cite{carlberg_2013}
which shows the shape of the stream in an N-body simulation, where $x$
is the radial direction and $y$ is the tangential direction along the
orbit. Projections of these curves onto the y axis give the density
along the stream. Although the curves are not labeled by their time,
the early density enhancement is visible from the curves which are
steep near $y=0$. In addition, projections of the saw-tooth shape in
that figure give the caustics described in this paper.

The results of this work can be used to shed light on the results of
N-body simulations of stream impacts in previous works. For example,
in \cite{carlberg_2012}, N-body simulations are used to determine the
density in the center of a gap from impacts with various mass subhalos
and impact parameters. This central density is then used to make cuts
on what mass subhalos and impact parameters would create observable
gaps. In \cite{carlberg_2012}, fits were made to the central density
as a function of mass and impact parameter but we now have an analytic
expression for this result, i.e. \eqref{eq:central_density_explicit},
which matches this behavior. However, we note that our analysis is for
flybys of Plummer spheres while \cite{carlberg_2012} uses spherical
Hernquist profiles \citep{hernquist_1990}.

Similarly, in \cite{yoon_etal_2011}, the authors show the results of
N-body simulations of stream impacts with NFW subhalos of varying mass
(Figure 6 of that work). The caustic timescale for their fiducial
simulation is 800 Myr so for the snapshots presented, the fiducial run
is well into the caustic phase. At the bottom panel of their Figure 6,
we see the effect of varying the mass. In the caustic phase, the gap
size is given by \eqref{eq:gapsizelate}, where it goes like
$\sqrt{M}$. Thus, if we increase the mass by a factor of 10, the gap
size should increase by a factor of 3, as seen in their figure. If we
decrease the mass by a factor of 10, the gap is now in the
intermediate phase but the gap will still be roughly 1/3 of the
fiducial gap size, as seen in their figure. This can be repeated for
the other panels to understand the quantitative trends seen.

\section{Conclusion} \label{sec:conclusion}

In this work, we have studied how gaps are created in tidal streams by
the close passage of a dark matter subhalo. We restricted our analysis
to streams on circular orbits which allowed us to tackle the problem
analytically. Our main results can be summarised as follows.

\begin{itemize}

\item We provide a parametric expression for the stream density (\eqref{eq:psioftfg} and \eqref{eq:genden}). We emphasize that
  this result allows one to determine the stream density at all times,
  for an arbitrary impact geometry and an arbitrary spherically
  symmetric potential. This can be used to make realistic matched
  filters for finding gaps in observed streams. We also note that this
  model can easily be extended to different subhalo profiles by
  computing the velocity kicks and the parametric function
  numerically.

\item We confirm that gap formation in tidal streams is a runaway
  process which can lead to dramatic density reduction across tens of
  degrees on the sky. However, as we show explicitly for the first
  time, the orbital perturbation inflicted by the subhalo depends on
  the shape of the effective potential around the impact
  point. Therefore, the rate of gap growth depends strongly on the
  mass distribution in the host galaxy: in extreme cases, e.g. in a
  spherical harmonic oscillator potential, the gaps will not develop at
  all.

\item We discover that the evolution of gaps in tidal streams proceeds
  in three distinct phases. First, there is a \textit{compression
    phase} since the subhalo pulls stream particles towards the point
  of closest approach. These kicks change the orbital period of each
  stream particle leading to the \textit{expansion phase}, which
  causes the compression to reverse, and eventually leading to the
  creation of a gap. Due to the change in the orbital period, stream
  particles which received large kicks will eventually pass those
  which received no kick, leading to the \textit{caustic phase} with
  caustics (particle pile-ups) on either side of the gap. We predict
  therefore four caustics altogether, each pair with a different
  behaviour as a function of time.

\item Our analytic model allows us to make quantitative predictions for
  each phase. During all phases, we have an expression for the central
  density of the gap. During the expansion phase, we have expressions
  for the gap size and the density in the peaks around the gap. During
  the caustic phase, we have expressions for the gap size, relative
  strength of the caustics, distance between the caustics, and width
  of the caustics.

\item Contrary to previous work, we unravel an important change in the
  gap growth at late times. Stream gaps stop growing as fast as $t$
  and switches to a slower rate proportional to $\sqrt{t}$ as the
  \textit{expansion phase} evolves into the \textit{caustic phase}.

\item In addition to the secular behavior described above, we
  demonstrate that the gap properties oscillate during all three
  phases due to epicyclic motion. These oscillations will become yet
  more pronounced for streams on eccentric orbits and, unfortunately,
  are bound to muddle any inference based on the gap properties.

\item We verified the analytical model with N-body numerical
  experiments. These include a set of idealized simulations with
  stream particles on circular orbits and found an almost perfect match with
  the gap profile, as well as the gap size, central density, and peak
  density. In addition, we compared the model to an N-body simulation
  where a globular cluster on a circular orbit is disrupted to create
  a realistic stream.  In this case, again the model describes the gap
  properties rather well, with a slight mismatch at late times, likely
  due to the spread in energy and angular momentum in the stream.

\item Finally, we take advantage of the analytic model to see how
  observations of gap profiles can be used to constrain the dark
  matter subhalo properties. When considering a single gap, we found a
  large degeneracy between the subhalo properties, the gap properties,
  the host potential and the epoch of observation. Even in the best
  case scenario when the entire gap profile can be matched, it is not
  possible to uniquely infer the properties of the subhalo and the
  geometry of the flyby. 

\end{itemize}

Let us stress once again that this qualitative picture outlined above
is quite general and will hold for other dark matter subhalo profiles,
non-circular orbits, and streams with a realistic distribution of
energy and angular momentum. The analytic expressions presented in
this work also allow us to quantitatively understand the trends seen
in N-body simulations of stream disruptions with varying impactors
\citep[i.e.][]{yoon_etal_2011,carlberg_2012}. While our study has
uncovered many degeneracies and complications inherent in the stream
spatter analysis, we have built a solid framework which can be used to
infer dark matter subhalo properties from tidal stream gaps.

Lastly, we have made two movies to showcase the different phases of gaps described in this work. The first movie shows the gap produced in the realistic simulation described in \Secref{sec:realistic_sim} and can be found \href{http://youtu.be/MXfKmnARBNM}{here}\footnote{Available through MNRAS and \href{http://youtu.be/MXfKmnARBNM}{http://youtu.be/MXfKmnARBNM}}. The second movie shows a gap produced using the same setup but with a smaller subhalo with $M=10^7 M_\odot$ and $r_s = 125$ pc and can be found \href{http://youtu.be/p0kqH5l0x3M}{here}\footnote{Available through MNRAS and \href{http://youtu.be/p0kqH5l0x3M}{http://youtu.be/p0kqH5l0x3M}}.

\section*{Acknowledgements}

We thank the anonymous referee for a helpful and thorough report. We thank the Streams group at Cambridge for stimulating discussions and in particular we thank Wyn Evans for detailed comments on this manuscript, as well as Sergey Koposov and
Jason Sanders for useful discussions. In addition, DE thanks Kerbal Space Program for
invaluable intuition gained from practicing orbital maneuvers. The
research leading to these results has received funding from the
European Research Council under the European Union's Seventh Framework
Programme (FP/2007-2013)/ERC Grant Agreement no. 308024.

\bibliographystyle{mn2e_long}
\bibliography{citations_gg}

\end{document}